# Effect of Precursors and Radiation on Soot Formation in Turbulent Diffusion Flame

Manedhar Reddy B[1], Ashoke De[1], Rakesh Yadav[2]

[1]Department of Aerospace Engineering, Indian Institute of Technology, Kanpur, India-208016
[2]Ansys Fluent India Pvt. Ltd., Pune, India-411057

**ABSTRACT**

Soot formation in 'Delft flame III', a pilot stabilized turbulent diffusion flame burning natural gas/air, is investigated using ANSYS FLUENT by considering two different approaches for soot inception. In the first approach soot inception is based on the formation rate of acetylene, while the second approach considers the formation rate of two and three-ringed aromatics to describe the soot inception [1]. Transport equations are solved for soot mass fraction and radical nuclei concentration to describe inception, coagulation, surface growth, and oxidation processes. The turbulent-chemistry interactions and soot precursors are described by the steady laminar flamelet model (SLFM). Two chemical mechanisms GRI 3.0 [2] and POLIMI [3] are used to represent the effect of species concentration on soot formation. The radiative properties of the medium are included based on the non-gray modeling approach by considering four factious gases; the weighted sum of gray gas (WSGGM) approach is used to model the absorption coefficient. The effect of soot on radiative transfer is modeled in terms of effective absorption coefficient of the medium. A beta probability density function (β-PDF) in terms of normalized temperature is used to describe the effect of turbulence on soot formation. The results clearly elucidate the strong effect of radiation and species concentration on soot volume fraction predictions. Due to increase in radiative heat loss with soot, flame temperature decreases slightly. The inclusion of ethylene has less synergic effect than that of both benzene and ethylene. Both cases have less impact on the nucleation of soot. The increase in soot volume fraction with soot-turbulence interaction is in consistence with the DNS predictions.

# NOMENCLATURE

$a_{\epsilon,i}$     emissivity weighting factor

$a_s$     characteristic strain rate

$b^*_{nuc}$     normalized radical nuclei concentration

$c_{p,i}$     $i^{th}$ species specific

$k_i$     absorption coefficient of gas

$k_{soot}$     absorption coefficient of soot

$f$     mixture fraction

$p$     sum of the partial pressures of all absorbing gases

$R^*_{nuc}$     normalized rate of nuclei generation

$R_{soot}$     rate of soot formation

$T$     Temperature

$Y_i$     species mass fraction

$Y_{soot}$     soot mass fraction

$\kappa_i$     absorption coefficient of $i^{th}$ gray gas

$\rho$     Density

$\sigma_{soot}$     turbulent Prandtl number of soot transport

$\phi$     representative scalar

$\chi$   scalar dissipation rate

$\chi_{st}$   scalar dissipation rate at $f = f_{st}$

$erfc^{-1}$   inverse complementary error function

## 1. INTRODUCTION

Soot formation and oxidation phenomenology mechanisms have undergone several strides towards making accurate quantitative estimates in combustion systems. There is considerable advance in the predictive capabilities, but complete understanding of the physics and chemistry still persists. To improve the understanding of soot formation detailed modeling is required in order to explain the production of Polycyclic Aromatic Hydrocarbons (PAH), the building blocks of soot. In addition, detailed modeling can be used to provide details regarding the number, size and shape of the particles. Soot formation involves highly non-linear physical and chemical processes. The evolution process of soot has been reviewed by Haynes and Wagner [4], Bockhorn [5], Kennedy [6] and have been classified into four major sub processes. The formation of PAH, conglomeration of PAH, surface reaction of particles (growth and oxidation) and agglomeration of particles. The accurate modeling of this sub processes is required for the accurate estimation of soot evolution. Modeling soot formation in turbulent diffusion flames is particularly a challenging task due to the small scale interactions between turbulence, chemistry and soot particle dynamics. Description of soot precursors such as acetylene and PAH by the combustion model imposes additional constraints as it involves the resolution of a large number of reactions involving stiff chemistry. Considering radiative heat transfer further increases the complexity, as it affects the kinetic rate associated with soot precursors and creates a two way coupling between combustion and soot models. Soot has high temporal and spatial intermittency due to non-linear interactions between turbulence, molecular transport and chemistry.

The soot nucleation and growth was considered as a first-order function of acetylene concentrations by Brooks and Moss [7], the model was applied to a pilot stabilized turbulent diffusion methane/air flame and reasonable good predictions were obtained. Kronenburg et al. [8] investigated the effect of turbulence-chemistry interaction on the same flame by using conditional moment closure instead of the extended laminar flamelet model as used in the work of Brooks and Moss. A comprehensive RANS study was performed by Pitsch et al. [9], in this approach a moment method was used to calculate soot evolution. The results were in good agreement with the experimental measurements. The H-Abstraction-Carbon-Addition (HACA) mechanism, is based on the concentration of acetylene only, has sometimes under estimated the soot formation rate, indicating the importance of odd-carbon chemistries like, propargyl ($C_3H_3$), cyclopentadienyl ($C_5H_5$) and benzyl ($C_7H_7$) on the PAH and soot formation rate. Propargyl recombination is considered one of the dominant paths in the formation of benzene. The recent LES study by Mueller et al. [10] on soot formation in Delft Flame III indicated the presence of significant uncertainty in the PAH chemistry. The soot volume fraction was found to be sensitive to the sub-filter variance and showed qualitative agreement with the experimental intermittency. In recent DNS studies, the interaction of turbulence and soot was investigated in terms of transport properties and the evolution of PAH was also studied. The role of transport properties was emphasized by Lignell et al. [11], the work highlighted that the probability of soot moving toward or away from the flame had the same probability. Bisetti et al. [12] used a soot inception model based on PAH instead of an acetylene based model as used by Lignell et al. [11]. It was observed that soot was sensitive to the scalar dissipation rate, this result in intermittency of soot and the slow chemistry results in significant unsteady kinetic effects.

The resolution of chemistry involving soot formation is difficult as it involves the resolution of large number of stiff reactions. Hence mixture fraction based approaches are generally applied. The objective of the present study are (i) to investigate the effect of species concentration and radiation on soot evolution; (ii) to estimate the synergy effect of ethylene and benzene on soot nucleation and surface growth, in addition study the influence of PAH based inception model on soot evolution; and (iii) to study the effect of soot-turbulence interaction on soot evolution process. Different approaches for prediction the OH radical concentration has been examined and their effect on soot formation has been discussed. The results are compared with the experimental measurements of Qamar et al. [13] in order to shed light on the understanding of soot formation in this flame.

## 2. NUMERICAL DETAILS

### 2.1 Governing Equations

The Favre averaged governing equations of mass, momentum, energy and turbulence are solved which has the following general form as:

$$\frac{\partial}{\partial t}\rho\tilde{\phi} + \frac{\partial}{\partial x_j}\rho u_j\tilde{\phi} = \frac{\partial}{\partial x_j}\rho D \frac{\partial\tilde{\phi}}{\partial x_j} + \langle S_\phi \rangle \tag{1}$$

Where the Favre averaged velocity in the j$^{th}$ coordinate direction is represented by $u_j$, and $\rho$ is the mean density. The Favre averaged scalar in the turbulent flow filed is given by $\tilde{\phi}$ and $\langle S_\phi \rangle$ represents the mean scalar term of the scalar and $D$ is the coefficient of scalar diffusion. The turbulence scale information is provided by the RSM model.

### 2.2 Turbulence-chemistry Interactions

The laminar flamelet model considers the turbulent flame as an ensemble of laminar and one dimensional local structures [14]. The flame surface is defined as an iso-surface of the mixture fraction within the turbulent flow field. Counter flow configuration of laminar diffusion flame is used to represent the thin reactive-diffusive layers in the turbulent flow field. The flame equations can be transformed from physical to mixture fraction space to represent the reactive-diffusive layer and are given by:

$$\frac{\partial Y_i}{\partial t} = \frac{1}{2}\chi\rho\frac{\partial^2 Y_i}{\partial f^2} + S_i \tag{2}$$

$$\rho\frac{\partial T}{\partial t} = \frac{1}{2}\rho\chi\frac{\partial^2 T}{\partial f^2} - \frac{1}{C_p}\sum_i H_i S_i + \frac{1}{2C_p}\rho\chi\left[\frac{\partial C_p}{\partial f} + \sum_i C_{p,i}\frac{\partial Y_i}{\partial f}\right]\frac{\partial T}{\partial f} \tag{3}$$

The mixture fraction and scalar dissipation can be used to map species mass fraction and temperature from mixture fraction space to physical space. The scalar dissipation rate quantifies the deviation from equilibrium and is defined as:

$$\chi = 2D\left|\frac{\partial f}{\partial x_j}\right|^2 \tag{4}$$

The scalar dissipation rate varies along the flamelet and is modeled as Eq. [5] with a small stoichiometric mixture fraction [15] and variable density as:

$$\chi_{st} = \frac{a}{4\pi}\frac{3(\sqrt{\rho_\infty/\rho_{st}}+1)^2}{(2\sqrt{\rho_\infty/\rho_{st}}+1)}\{-2[erfc^{-1}(2f_{st})]^2\} \tag{5}$$

The ensemble of diffusion flamelets are used to represent the turbulent flame brush. The Favre averaged species mass fraction and temperature for a turbulent flame can be determined as:

$$\tilde{\phi} = \iint \phi(f,\chi_{st})p(f,\chi_{st})dfd\chi_{st} \tag{6}$$

Where a presumed $\beta-\text{PDF}$, used to define the probability of the mixture fraction. The temperature and mean density have an extra dimension of mean enthalpy $\tilde{H}$ to consider the non-adiabatic steady diffusion flamelets. The species mass fraction is assumed to have negligible effect by the heat loss or gain by the system. The evolution of mixture fraction in the physical space is represented by the transport equation of $f$ and $f''$ is given as:

$$\frac{\partial}{\partial t}(\rho\tilde{f}) + \frac{\partial}{\partial x_k}\rho u_k \tilde{f} = \frac{\partial}{\partial x_j}\left(\frac{\mu_t}{\sigma_t}\frac{\partial \tilde{f}}{\partial x_k}\right) \tag{7}$$

$$\frac{\partial}{\partial t}(\rho\tilde{f}^{"2}) + \frac{\partial}{\partial x_k}\rho u_k \tilde{f}^{"2} = \frac{\partial}{\partial x_j}\left(\frac{\mu_t}{\sigma_t}\frac{\partial \tilde{f}^{"2}}{\partial x_k}\right) + C_g\mu_t\left(\frac{\partial \tilde{f}^{"2}}{\partial x_k}\right)^2 - C_d\rho\frac{\varepsilon}{k}\tilde{f}^{"2} \tag{8}$$

Where $\sigma_t$, $C_g$ and $C_d$ are 0.85, 2.86 and 2.0 respectively.

## 2.3 Radiation Modeling

The WSGG model, four fictions gases are used to represent the non-gray medium. The details of the approach can be found in Rakesh et al. [16]. Total emissivity over a distance $s$ is given by Eq. (9).

$$\epsilon = \sum_{i=0}^{I} a_{\epsilon,i}(T)(1 - e^{\kappa_i ps}) \tag{9}$$

The medium is optically thick and radiative transfer equation can be approximated into a truncated series expansion in spherical harmonics (P1 approximation) [17] given by Eq.. (10)

$$-\nabla \cdot \left(\frac{1}{3(K_{av} - K_{sv}) - Ak_{sv}}\nabla G_v\right) = K_{av}(K_{bv} - G_v) \tag{10}$$

Where the right hand side of Eq. 10 represents the radiative source added to the energy equation. $K_{av}$ represents the absorption coefficient of the product gasses and soot.

## 2.4 Soot Modeling

In the Moss-Brookes model two transport equations in terms of soot volume fraction and nuclei concentration are used to describe the evolution of soot, they are described as:

$$\frac{\partial \rho Y_{soot}}{\partial t} + \nabla \cdot (\rho \bar{v})Y_{soot} = \nabla \cdot (\frac{\mu_t}{\sigma_{soot}}\nabla Y_{soot}) + \frac{dM}{dt} \tag{11}$$

$$\frac{\partial \rho b^*_{nuc}}{\partial t} + \nabla \cdot (\rho \bar{v})b^*_{nuc} = \nabla \cdot (\frac{\mu_t}{\sigma_{nuc}}\nabla b^*_{nuc}) + \frac{1}{N_{norm}}\frac{dN}{dt} \tag{12}$$

The soot particle number density (N) and mass density (M) are used in the Brookes and Moss [7] model to define the soot formation in turbulent diffusion flames. The soot particles instantaneous production is defined by the gas phase particles nucleation and coagulation, given by Eq. (13). The net soot production is the source term for the soot mass concentration and is calculated using Eq. (14).

$$\underbrace{\frac{dN}{dt} = C_\alpha \cdot N_A \left(\frac{X_{prec}P}{RT}\right) \exp\left\{-\frac{T_\alpha}{T}\right\}}_{Nucleation} - \underbrace{C_\beta \cdot \left(\frac{24RT}{\rho_{soot}N_A}\right)^{1/2} d_p^{1/2} N^2}_{Coagulation} \tag{13}$$

$$\frac{dM}{dt} = \underbrace{C_\alpha \cdot M_P \cdot \left(\frac{X_{prec}P}{RT}\right) \exp\left\{-\frac{T_\alpha}{T}\right\}}_{Nucleation} + \underbrace{C_\gamma \cdot \left(\frac{X_{sgs}P}{RT}\right) \exp\left\{-\frac{T_\gamma}{T}\right\} \left[(\pi N)^{1/3} \left(\frac{6M}{\rho_{soot}}\right)^{2/3}\right]}_{Surface-Growth}$$
$$- \underbrace{\left[C_{oxid} \cdot \sqrt{T} (\pi N)^{1/3} \left(\frac{6M}{\rho_{soot}}\right)^{2/3} \cdot \left(C_{\omega,1} \cdot \eta_{coll} \left(\frac{X_{OH}P}{RT}\right) \exp\left\{-\frac{T_{\omega,1}}{T}\right\} + C_{\omega,2} \cdot \left(\frac{X_{O_2}P}{RT}\right) \exp\left\{-\frac{T_{\omega,2}}{T}\right\}\right)\right]}_{Oxidation} \tag{14}$$

2.4.1 Nucleation and Coagulation

The nucleation of PAH molecules result in the formation of the first soot particles. Two models are used in the present study to examine the effect of inception rate on soot formation. The first model considers acetylene as the primary soot precursor and calculates the soot inception rate as a linear function of acetylene concentration, as given in Eq. (13). The constants of the model are, $N_A$ (=6.022045×10$^{26}$ kmol$^{-1}$) is the Avogadro number, $X_{prec}$ (C$_2$H$_2$ for the Moss-Brookes model) is the concentration of soot precursor and activation temperature of soot inception process is 21100 K; was proposed by Lindstedt [18]. In Eq. (13), $d_p$ is the soot particle mean diameter and $\rho_{soot}$ is the mass density of soot; fixed as 1800 kg/m$^3$ [7].

Hall et al. [1] extended the soot inception rate model, by considering the formation of two and three ringed aromatic in the inception rate as given in Eq. (15).

$$\left(\frac{dN}{dt}\right)_{Inc.} = 8 \cdot C_{\alpha,1} \cdot \frac{N_A}{M_P} \left[\rho^2 \left(\frac{Y_{C_2H_2}}{W_{C_2H_2}}\right)^2 \frac{Y_{C_6H_5} W_{H_2}}{W_{C_6H_5} Y_{H_2}}\right] \exp\left\{-\frac{T_{\alpha,1}}{T}\right\}$$
$$+ 8 \cdot C_{\alpha,2} \cdot \frac{N_A}{M_P} \left[\rho^2 \left(\frac{Y_{C_2H_2}}{W_{C_2H_2}}\right)^2 \frac{Y_{C_6H_6} Y_{C_6H_5} W_{H_2}}{W_{C_6H_6} W_{C_6H_5} Y_{H_2}}\right] \exp\left\{-\frac{T_{\alpha,2}}{T}\right\}$$
(15)

Where $C_{\alpha,1} = 127 \times 10^{8.88}$, $C_{\alpha,2} = 178 \times 10^{9.50}$, $T_{\alpha,1} = 4378$ K and $T_{\alpha,2} = 6390$ K as determined by Hall et al. [1]. The mass of soot incipient particles were assumed to be 1200 kg/kmol (equivalent to 100 carbon atoms, 12 carbon atoms were used by Brookes and Moss [7]. The mass density of soot was considered to be 2000 kg/m$^3$, slightly different from the value used by Brookes and Moss [7]. The rate of change of soot particle number density and mass density are related as:

$$\left(\frac{dM}{dt}\right)_{Inc.} = \frac{M_P}{N_A} \left(\frac{dN}{dt}\right)_{Inc.}$$
(16)

Collision of soot particles results in coagulation. The collision frequency is dependent on the size of the particles and mean free path of the gas phase particles. The soot particles are assumed to be spherical and smaller than the mean free path of the gas. The collision frequency defined by Puri et al. [19] is used in the present study.

### 2.4.2 Surface Growth

The addition of carbon atoms on the surface of soot particles due to H-Abstraction-Carbon-Addition (HACA) [20] with gas phase particles result in surface growth. The effect of acetylene on growth was described by Frenklach et al. [21] and Harris et al. [22]. Therefore, growth can be parameterized in terms of acetylene concentration and active area density on the soot surface. The active density is a function of the mono-dispersed particles surface area. The activation temperature for the surface growth was predicted to be 12,100 K by Leung et al. [23].

### 2.4.3 Oxidation

In present study both OH and $O_2$ are considered in the oxidation of soot. Initial studies were [24] carried considered only OH radical as the oxidizing agent. The effect of O2 on soot oxidation was studied by Lee et al. [25]. The collisional efficiency parameter and surface kinetics were used as the limiting factors in determining the rate of oxidation due to molecular oxygen. The effect of diffusion on soot oxidation is neglected, as the particles are small.

### 2.5 **Determination of OH Radical Concentration**:

In the Present study an equilibrium approach is used to determine the OH concentration. The over prediction of flame temperature results in overshoot of oxidizing radicals; resulting in under prediction of soot. Thus the concentration of OH is calculated as [26-27] Eq. (17).

$$[OH] = 2.129 \times 10^2 T^{-0.57} e^{-4595/T} [O]^{1/2} [H2O]^{1/2} \quad \text{gmol/m}^3 \tag{17}$$

The concentration of O-radical is determined in three approaches as stated below.

#### 2.5.1 Equilibrium Approach

The equilibrium concentration of O-radical is given by Westenberg [28] as:

$$[O] = 3.97 \times 10^5 T^{-1/2} e^{-31090/T} [O_2]^{1/2} \quad \text{gmol/m}^3 \tag{18}$$

#### 2.5.2 Partial Equilibrium Approach

In this approach the third body reaction in $O_2$ dissociation is also considered and O-radical concentration is given as:

$$[O] = 36.64 T^{1/2} e^{-27123/T} [O_2]^{1/2} \quad \text{gmol/m}^3 \tag{19}$$

#### 2.5.3 Instantaneous

In this approach the species concentration of O-radical is the estimated from the combustion model.

## 2.6 Soot-Radiation and Soot-Turbulence Interactions:

The absorption coefficient of the medium is greatly affected with the presence of soot. The effect of soot on the radiative heat transfer is included using an effective soot absorption. The effective absorption coefficient of the summation of the absorption coefficient of soot and gas:

$$\kappa_{eq} = \kappa_i + \kappa_{soot} \tag{20}$$

The absorption coefficient due to soot particulates is given as:

$$\kappa_{soot} = b_1 \rho_{soot} \left[1 + b_T (T - 2000)\right] \tag{21}$$

where $b_1 = 1232.4 \, m^2/Kg$ and $b_T = 4.8 \times 10^{-4} \, K^{-1}$ are given by Taylor et al. [29] and Smith et al. [30].

The interaction between soot, temperature and species concentrations are highly non-linear. Hence, inclusion of temperature and species composition fluctuations will have a significant effect on soot. The turbulence effects are included by considering a single variable PDF in terms of temperature to describe the temporal variation.

$$\tilde{S}_{soot} = \int \rho S_{soot}(T) P(T) dT \tag{22}$$

The PDF is constructed from the independent variables obtained from the solution of soot transport equations. An extra transport equation is solved to calculate temperature variance. The maximum and minimum temperatures in the flame are used as the limit of integration. The maximum flow filed temperature obtained from the combustion solution is used as the upper limit and minimum flow field temperature is used as lower limit of integration.

## 3. Description of Test Case

Delft Flame III, a pilot stabilized non-premixed natural gas flame is used in the present study. It is one of the flame used in Turbulent Non-premixed Flame (TNF) workshops [31]. The details of the burner are provided by Peeters et al. [32]. The fuel jet has a diameter of 6 mm and is surrounded by a rim with an external diameter of 15 mm. The rim also acts as a bluff body creating a recirculation region for hot gases which stabilize the flame, twelve pilot flames of 0.5 mm diameter emanate from the rim stabilizing the flame. The power of the hot equilibrated mixture entering the domain is ~200 W and accounts for less than ~1% the power of jet. The rim is followed by a concentric circular inlet of internal diameter of 15 mm and external diameter 45 mm, which acts as the inlet for the primary air. The concentric face of the outer face acts as the secondary air inlet.

The velocity measurements were performed by Stroomer [33] using Laser Doppler Anemometry (LDA). Raman-Rayleigh-LIF (Laser Induced Fluorescence) technique was used by Nooren et al. [34] to measure the thermochemical scalars. The measurements were confined to the region close to burner and where soot is not present. Laser Induced Incandescence (LII) was used by Qamar et al. [13] to measure the sot volume fraction in the downstream. As the measurements were made in different places, the commercial natural gas composition was different. Nooren et al. [34] and Qamar et al. [13] diluted the natural gas with $N_2$ to match the adiabatic flame temperature of Stroomer [33].

## 4. Numerical Details

In the present work, ANSYS FLUENT 15.0 [35] along with the UDF for the radiative heat transfer model is used to perform all the calculations. A second order unwinding scheme is used to discretize the convective fluxes. The pressure-velocity coupling is implemented using the SIMPLE algorithm. As the flame is significantly affected by the predictions in velocity profiles, Reynolds stress (RSM) model [36] is used model the turbulence. The steady laminar flamelet model (SLFM) [37] is used to model the turbulence chemistry interactions for the gas-phase chemistry. The WSGG method is used to model the radiative heat transfer. The non-gray behaviour is represented using four-fictitious gray gases, whose weight functions and absorption coefficients are calculated from Smiths et al. [30] as a function of $H_2O$ and $CO_2$ partial pressures. The details of the approach can be found in Rakesh et al. [16].

4.1 Modeling Details

As the flow is symmetric, an axisymmetric 2D grid is used in the present computational work. The diameter of the pilot fuel inlet is determined from the equivalent area of the 12 holes on the rim. For the present study a computational domain of 250D x 50D, where D is the diameter of the fuel jet (Fig. 2) has been considered. Initially, the grid independence study is performed with two different grids of 400 X 300 and 200 X 150 points and the predictions of the two grids are shown in Fig. 3; as the predictions using both the grids are in good agreement with each other, the coarse grid (200 X 150) is chosen for rest of the computations, where 13, 12 and 6 grid points are used to resolve the fuel inlet, pilot fuel, pilot wall rim. The mass fractions of the fuel are adjusted such that it consists of 81% $CH_4$, 4% $C_2H_6$ and 15% $N_2$ by volume. The calorific value remains same as the fuel used by Stroomer [33] and the stoichiometric mixture fraction is 0.07. GRI 3.0 mechanism containing 52 species and 325 reactions [2], and POLIMI mechanism containing 82 species and 1450 reactions [3] are used to represent the chemistry.

4.2 Boundary Conditions

The experimental boundary conditions are given in Table 1. To avoid spreading of flow in the upstream direction, the boundary conditions at the inlet are specified with the velocity profiles at X = 3 mm. In order to maintain the experimental mass flow rate, an iterative approach is used to correct the magnitude of velocity profiles to match the mass flow rate at the inlet, while maintaining the shape of the profiles. This method was earlier used [32, 38] with this burner. The experimental data at an axial distance of 3 mm from the jet exit is used as inlet condition for turbulent kinetic energy, turbulent dissipation and Reynolds stress components [32]. The modelling of the pilot flames which control the flame stabilization at the exit of the burner is one of the most challenging part of modeling this flam. The pilot flame inlet has been modified by increasing the mass flow rate as suggested by Naud et al. [39]. Mixture fraction is used to represent the higher hydrocarbons in the pilot flame, the value is 0.115. As suggested by Naud et al. [39] and Rakesh et al. [38, 40] the mean pilot velocity of 8.28 m/sec is used in our calculations.

**5. Results and Discussion**

In this section, the flamelet computations are presented and compared with experimental measurements. The study also includes the effect of radiative heat transfer on the flame temperature. Then, the soot results are compared with the experimental measurements; followed by a qualitative study of the soot formation. The soot predictions are also compared to a recent LES study using a hybrid method of moment's method for soot modeling. The effect of radiation and turbulence on soot has been comprehensively studied.

5.1 Velocity

The radial profiles of mean and RMS axial velocity are shown in Fig. 4. The agreement between the calculated and experimental measurements is found to be excellent. The mean axial velocity profiles show slight under prediction along the downstream direction and within the uncertainty of experimental measurements [33]. The increase in diffusion due to over predicted mean temperature results in the under-estimation of velocity along the downstream positions. The under prediction in the downstream positions is less when compared to the previous RANS [41] and LES [42] studies. The LES predictions of Mueller et al. [10] using the RFPV model are similar to the present predictions. The current model predictions differ from the predictions of Rakesh et al. [38], mainly due to the difference in pilot modeling approach and joint composed PDF model was used as opposed to steady flamelet in the present study.

A similar comment can be made on the RMS and turbulent kinetic energy predictions. The RMS are also similar to the LES predictions of Mueller et al. [10] and better than the predictions of Ayache and Mastorakos [42]. This agreement can be attributed to the velocity and turbulence boundary conditions used in the present study. The under prediction near the burner by Ayache and Mastorakos [42] might be due to usage of bulk velocities for fuel and air coflow; and zero turbulence boundary conditions. The two lines in Fig 4 represent the simulations using two different chemical mechanisms. The POLIMI mechanism might give a very slight improvement in the predictions of mean and RMS axial velocity than the GRI mechanism, but the velocity and scalars are not strongly affected in the upstream direction.

5.2 Mixture fraction

Fig. 5 shows the radial profiles of mean and RMS mixture fraction. Similar to the velocity predictions the calculated and experimental measurements are in good agreement. The potential experimental uncertainty in the experimental measurement of mixture fraction was reported to be typically less than 9% for lean mixtures and was typically 5% uncertainty for rich mixtures and a maximum of 20% by Nooren et al. [34]. The discrepancies observed are, the spreading rate is slightly higher near the burner and the mixture fraction is over predicted in the centerline along the downstream positions.

The over prediction of spreading rate can be linked to convection–diffusion equation, as it governs the evolution of the mean mixture fraction. The turbulent kinetic energy is over predicted along the radial direction near the burner, resulting in increase of spreading rate and over prediction of mean mixture fraction. The mean mixture fraction is over predicted from X=100 mm. The over prediction of centerline mixture fraction is due to excessive convection at the axis, due to over estimation of mean density because of the under estimation of mean velocity and temperature. The RMS mixture fraction predictions are better than the predictions of Ayache and Mastorakos [42] and are also similar to the LES predictions of Mueller et al. [10].

5.3 Temperature

The predicted radial profiles of mean and RMS temperature are reported in Fig. 6. The calculated mean temperature profiles are in reasonable agreement with the experimental measurements. The mean temperature profile shape is in good agreement near the axis but the radial peak temperature is slightly over predicted. This broadening of the temperature profiles is consistent with the corresponding mixture fraction profile. Despite the over prediction of mean mixture fraction about the centerline which should result in a lower temperature prediction about the centerline, the temperature is in good agreement with the experimental predictions. The failure to estimate local flame extinction results in the over prediction of mixture fraction, leading to good temperature estimate along the centerline. In the previous RANS studies of Merci et al. [41] and Habibi et al. [43] using the adiabatic steady flamelet model, this failure of estimating of local flame extinction resulted in over prediction of temperature globally. The LES study of Mueller et al. [10] captures the correct amount of local extinction resulting in better temperature predictions. A compelling statement about the accuracy of flame length prediction cannot be made since the experimental data is only available for the region close to the burner. The RMS temperature profiles are significantly under-predicted along the radial direction compared to the experimental measurements due to ignoring of local flame extinction. The predictions of Mueller et al. [10] are better compared to present study due to the inclusion of sub-filter scale scalar dissipation.

5.3.1 Inclusion of radiative heat transfer

Rakesh et al. [38] using joint composition PDF and Habibi et al. [43] using flamelet model, investigated the effect of radiation for this flame. They have observed that with the inclusion of radiative heat transfer the temperature reduced by ~100K at downstream positions as ~ 250 mm. Fig. 7 shows the axial profile of mean temperature with different radiation approaches. The centerline temperature drops by ~100K to ~2000K with the inclusion of gray and the peak temperature shifts by ~8D in the upstream direction. The temperature drops further by ~300K to ~1700K and shifted ~15D in the upstream direction with the non gray model. The location of radial maximum temperature was not affected by radiation. The global maximum temperature reduced from 2090K to 2040K with the gray radiation model and non-gray model gives a drop of ~60K with peak value of 1980K. The global maximum temperature shifted from the centerline in the radial direction by ~2D with gray radiation model and ~4D with the non-gray radiation model.

5.4 Species Mass Fraction

Radial profiles of species mass fractions at different axial locations using two chemical mechanisms are shown in Fig. 8. So far (Figs. 4-6), there is no substantial differences are observed for different chemical mechanisms; however, the species profiles are apparently affected by the kinetics. The predicted species mass fractions are found to be in good agreement with the experimental measurements and are consistent with the discrepancies observed with velocity and scalars. OH is under predicted about the center line and over predicted about the radial maximum location. CO and $H_2$ are over prediction about the centerline is consistent with the corresponding mixture fraction and temperature profile predictions. A similar comment can be made on the prediction of $H_2O$. Usage of detailed

POLIMI mechanism results slight reduction in the species prediction when compared to GRI 3.0 mechanism, there is better agreement with the experimental measurements. Considering the large experimental uncertainties as explained by Nooren et al. [34], the overall agreement of species mass fraction can be considered good.

The axial profile of OH mass fraction is shown in Fig. 7. The inclusion of radiation has no significant effect on the maximum mass fraction of OH, the location of maximum mass fraction is shifted upstream by ~3D with gray radiation and ~10D with the non-gray model with the GRI mechanism. A similar trend is observed with the POLIMI mechanism, ~5D with gray radiation and ~12D with the non-gray model. Since the experimental data is only available in the upstream region; no compelling statement can be made on the accuracy of computed OH mass fractions with different radiation approaches. The calculated OH equilibrium profile is slightly under predicted than that of OH calculated by the combustion model. The under prediction is due to the neglecting of radical overshoots.

5.5 Soot Predictions

The axial profile of centerline soot volume fraction is shown in Fig. 9 and compared with the experimental measurements of Qamar et al. [13] and the LES predictions of Mueller et al. [10]. Acetylene concentration is considered as the primary soot precursor and surface growth source. Equilibrium and par-equilibrium models are used to calculate the OH radical concentration, for the oxidation of soot. Fig. 9(a) shows the mean soot volume fraction without radiation, an over prediction by a factor of thirty is observed with the equilibrium model and a factor of fifteen with the partial equilibrium model using the GRI 3.0 chemical mechanism. The over prediction is reduced to a factor of ten with equilibrium model and five with the partial equilibrium model using the POLIMI mechanism. A similar trend has been observed with the inclusion of gray radiation as depicted in Fig. 9(b). With the equilibrium model the over prediction is found by a factor of twenty-five and seven with GRI and POLIMI mechanism, respectively. The over prediction is in the order of twelve and four with the partial equilibrium model. The predictions improve with the inclusion of the non-gray radiation model; soot volume fraction is over predicted by a factor of six and two with the equilibrium model. The predictions are reasonably good with the partial equilibrium model for both the GRI and POLIMI mechanism.

The effect of radiation on soot formation can be clearly elucidated form the over prediction of soot volume fraction without radiation; indicates the dependence on temperature and the need to account for the radiative heat transfer accurately. The location of maximum temperature about the centerline is not affected by the soot-radiation interactions. The rates of formation of PAH are substantially affected with the inclusion of radiation. The over prediction of soot volume fraction can be directly related with the over prediction of soot nucleation at high temperatures (in the absence of radiation). The balance of production and consumption in PAH are accurately captured, resulting in accurate predictions of maximum soot volume fraction location when compared to Mueller et al. [10] and Donde et al. [44]. The over prediction of soot volume fraction by Muller et al. [10] can be attributed to the usage of higher sub filter dissipation rate. The uncertainty in PAH mechanism used for formation and growth resulted in the formation of soot closer to the nozzle compared to experimental predictions. The over prediction can also be asserted to the consideration of nucleation as a major source of soot mass addition.

The source terms of soot mass concentration with gray and non-gray radiation models are shown in Figs. 10 and 11 respectively. It can be observed that the soot nucleation varies by an order of magnitude, indicating the strong effect of radiation on the production of PAH. It can also be inferred that the surface growth is also highly dependent on temperature. The reduction in soot volume fraction with the POLIMI mechanism can be attributed to the decline in acetylene concentration prediction as compared to the GRI mechanism. The overall PAH concentration depends on the chemical mechanism used for modeling combustion. The POLIMI mechanism gives an accurate prediction of the PAH concentration, compared to GRI 3.0 mechanism. The results expounded the effect of acetylene concentration on soot volume fraction and the need for accurate estimation of species mass fraction. In the LES study by Mueller et al. [10], there was major amount of soot was formed earlier than in the experiment. The discrepancy was attributed to the uncertainty in the PAH formation and growth kinetic mechanism. The present study is able to reasonable replicate the location of maximum soot volume fraction, while the over prediction of OH radical concentration results in rapid oxidation of soot. The soot volume fraction is under predicted by a factor of hundred when OH radical concentration is obtained from the combustion model, indicating the over prediction of OH-radical and the sensitivity of soot nucleation to the oxidizer radical. The inability of acetylene based models to account for the dependence of soot yield on local mixture fraction results in inaccurate prediction of soot volume fraction. The inclusion of higher order hydrocarbons and usage of PAH based inception mechanism is used to comprehend this effect.

5.5.1 Effect of Ethylene on nucleation and surface growth

The predicted soot volume fraction with the addition of ethylene in the computation of nucleation and surface growth source terms, using the non-gray radiation model is shown in Fig. 12. The inclusion of ethylene has a synergic effect on the soot volume fraction for both the mechanisms. The POLIMI mechanism predicts significant higher increase in soot volume fraction than the GRI mechanism with the addition of ethylene. This can be due to the higher ethylene prediction by the POLIMI mechanism. The partial mixing of oxygen with ethylene results in the generation of oxygen radicals that increase the soot formation, due to the formation of propargyl radicals as described by Hwang et al. [45]:

$$C_2H_2 + O = CO + CH_2$$
$$C_2H_2 + CH_2 = C_3H_3 + H$$
(23)

Fig. 13 shows the source terms of soot mass concentration with the addition of ethylene, and it can be observed that the addition of ethylene doesn't have considerable effect on nucleation about the centerline while the surface growth has increased by a factor of 1.5. It is also observed that the decay of nucleation occurs slowly with the flame height due to inexact knowledge of thermodynamics and kinetics of PAH. The over prediction of hydroxyl radicals and molecular oxygen results in the over estimation of soot oxidation near the flame sheet. The estimated hydroxyl radical and molecular oxygen concentrations are found to be less by factor of ten, compared to the predicted values. The under estimation of the predicted radical concentration results in the over prediction of soot volume fraction by the equilibrium and partial equilibrium approach. The results also elucidate the effect of ethylene on the HACA mechanism, the contribution of ethylene might be considered as the primary reason for the increase of surface growth source term. The distribution of soot surface growth in the mixture fraction space is accurately predicted, resulting in good agreement of the maximum soot volume fraction location prediction with experimental measurement.

The contour of soot mass concentration nucleation, surface growth and oxidation with and without the addition of ethylene are shown in Fig. 14. As explained earlier, nucleation is not considerable effected about the centerline with the addition of ethylene; however a slight increase can be observed about the reaction layer. Soot surface growth and oxidation have increase by three fold. As previously observed the surface growth is independent of the surface area of soot particles [46]. The inclusion of ethylene increases the utilization of active sites on the surface of particles. Hence the increase in surface growth can be explained as the change in dynamics of active sites resulting with the addition of ethylene. The increase in oxidation rate can also be explained in similar terms. The oxidation rate can be considered as a first order function of soot surface area. As the surface area increases with the inclusion of ethylene, the collision frequency of OH and $O_2$ increases. It can also be observed that the maximum oxidation rate occurs away from the centerline indicating the dominance of $O_2$ over the OH radical.

5.5.2 Effect of Benzene on nucleation and surface growth

Soot volume fraction with the addition of benzene and ethylene concentrations in the computation of nucleation and surface growth source terms are performed using the Moss-Brookes and Moss-Brookes-Hall models, as shown in Fig. 15. Soot is calculated only with POLIMI mechanism. The comparison shows that the Moss-Brookes-Hall model gives much higher prediction of soot than the Moss-Brookes model. Similar to ethylene, inclusion of benzene has a synergic effect on the soot volume fraction. The increase in soot volume fraction with PAH based inception model is due to the promotion of coagulation of smaller intact aromatic hydrocarbons to form large aromatics. The coagulation of nuclei has increased by a factor of 1000 with the Moss-Brookes-Hall model, indicating the effect of PAH on the growth mechanism. The soot volume fraction increases by a factor of three when OH concentration from the combustion model is considered, instead of calculated OH concentration using O radial concentration from the combustion model. This discrepancy might be due to the over prediction of calculated OH concentration. The O-Instantaneous approach gives better agreement with experimental measurements, indicating the need for accurate prediction of hydroxyl radical. The source terms of soot mass concentration using Moss-Brookes-Hall model are reported in Fig. 16. The over prediction of calculated OH doesn't exhibit substantial effect on the nucleation rate, but reduces the surface growth. It clearly indicates the effect of OH concentration on the soot growth rate, while the reduction is due to the oxidation of PAH before participating in surface reactions. This results in the shift in location of maximum soot volume fraction on the centreline by ~5D in the upstream direction. The radial profile of normalized soot volume fraction is shown in Fig. 17. The oxidation of soot already stated in the locations where the radial profiles are taken. The under prediction along the radial direction is due to rapid oxidation of soot in the presence of $O_2$ molecules. The under prediction along the radial direction increases with the decrease in OH concentration, this discrepancy is due to the substantial amount of oxidation by $O_2$.

6.6.2 Soot turbulence interaction

The effect of turbulence on soot formation is presented in Fig. 18. It can be observed that the soot volume fraction increases with the inclusion of turbulence. The temperature based PDF considers the over predicted mean temperature values as the limits of integration, resulting in over predicted average soot mass concentration rate. Accurate estimation of $\beta$ -PDF is required to capture the high temporal and spatial fluctuations. Fig 19 shows the axial profiles of soot mass concentration source terms, it can be inferred that the increase in soot surface growth is due to the increase in instantaneous temperature. The location of maximum soot volume is not significantly affected by turbulence. The over prediction of soot volume fraction can be directly related with the over estimation of surface growth. The balance of surface growth and oxidation are not accurately included, resulting in accurate predictions of maximum soot volume fraction when compared to Mueller et al. [10] and Donde et al. [44]. In LES study of Donde et al. [44], the soot-turbulence interaction increased the net soot production. The sensitivity of sub-filter dissipation rate has been attributed for this increase. The DNS study by Yoo and Im [47] for ethylene-air counter flow diffusion flame, found that particle growth reduces as soot spends less time at high temperature and the net soot produce increases due to increase in flame surface because of constrains. The contour of net soot production source term is shown in Fig. 20. The net soot production is confined mostly near the centreline with the O-instantaneous approach and spreads out with OH-instantaneous approach. The inclusion of turbulence increases the collision frequency in the shear layer. The net soot production shifts in the downstream direction with the OH-instantaneous approach. Due to formation of soot in the far downstream where turbulent fluctuations are minimum, net soot production is drastically effected by turbulence. Due to high intermittency, the steady RANS based approach is not able to the account for the complete temporal fluctuations. Thus further considerations are required in the formation of PDF to represent these fluctuations accurately.

## 7. CONCLUSIONS

In this work, soot formation in Delft III flame has been computed using the Steady Laminar Flamelet model in conjunction with two different soot approaches. The velocity and scalar predictions are in good agreement with experimental measurements. The velocity predictions tend to improve along the radial direction, whereas the mixture fraction is over predicted at the centerline location at downstream positions. The temperature is over predicted in shear layer and the RMS temperature is significantly under predicted. The radiative heat losses couldn't be accurately accounted by gray radiation. The centerline peak temperature reduced to ~1700K with the non-gray approach and the location also shifted in the upstream direction by ~15D. The maximum temperature is also reduced to 1980K and shifted ~4D above the centreline. Species prediction has a substantial impact on soot production. The predictions of soot are reduced by a factor of three with the POLIMI mechanism. The soot predictions with non-gray approach give a reasonable good match with experimental measurements. Due to over prediction of soot oxidizer radicals, soot volume fraction is substantially under predicted with O-instantaneous approach. The inclusion of ethylene has a synergic effect on soot volume fraction and expounds the effect of ethylene on HACA mechanism. Similar to ethylene, benzene also shows the synergic effect. Soot volume fraction increases by a factor of 100 with PAH based inception model, indicating the role of higher order PAH on the evolution of soot. Due to infrequent sooting event, temporal fluctuations are not accounted accurately which needs further investigation.

ACKNOWLEDGMENT

The authors would like to acknowledge the IITK computer center (www.iitk.ac.in/cc) for providing the resources to perform the computation work, data analysis and article preparation.

| | |
|---|---|
| $U_{fuel}$ | 21.9 m/sec |
| $T_{fuel}$ | 295K |
| $Re_{fuel}$ | 9700 |
| $U_{annulus}$ | 4.4 m/sec |
| $T_{annulus}$ | 295K |
| $Re_{annulus}$ | 8800 |
| $U_{coflow}$ | 0.4 m/sec |
| $T_{coflow}$ | 295K |

Table 1: Experimental data of delft III burner [32]

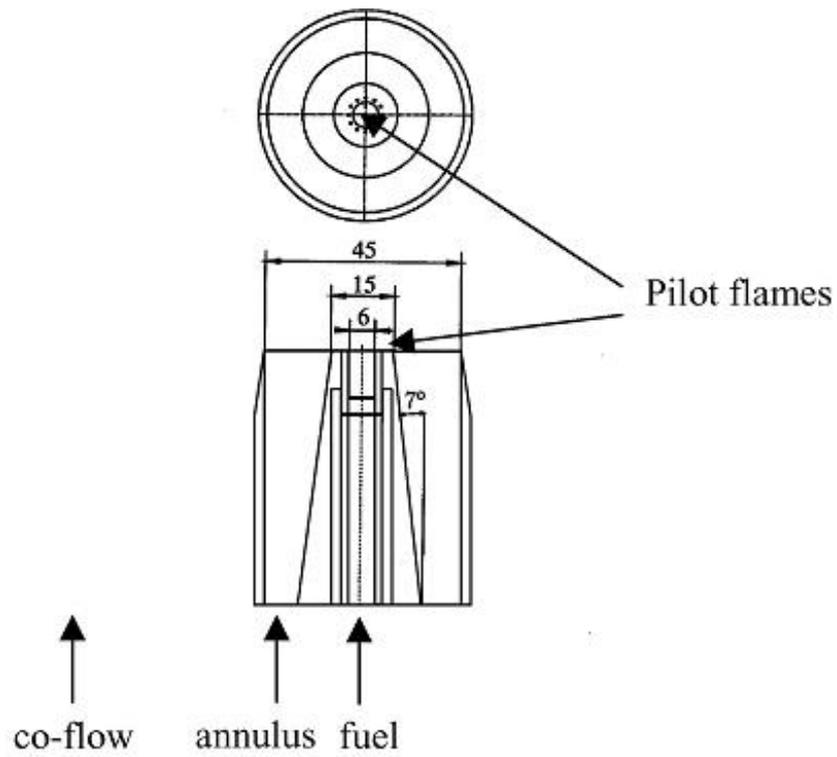

Figure 1: Top and cross view of axisymmetric non-premixed jet burner [32]

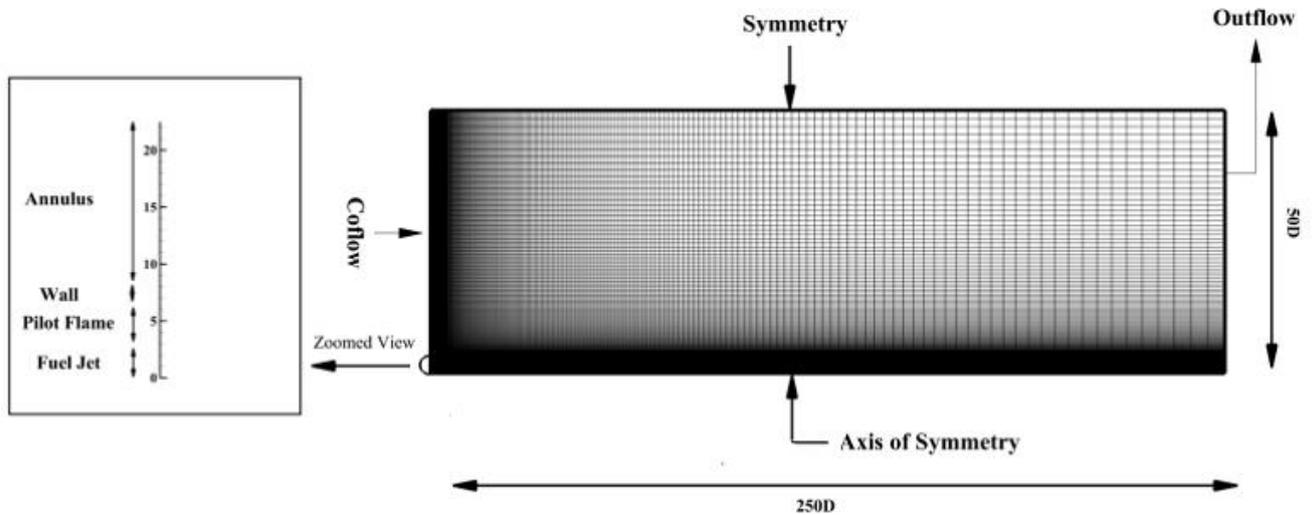

Figure 2: Illustrative view of the mesh used [D: diameter of fuel jet]

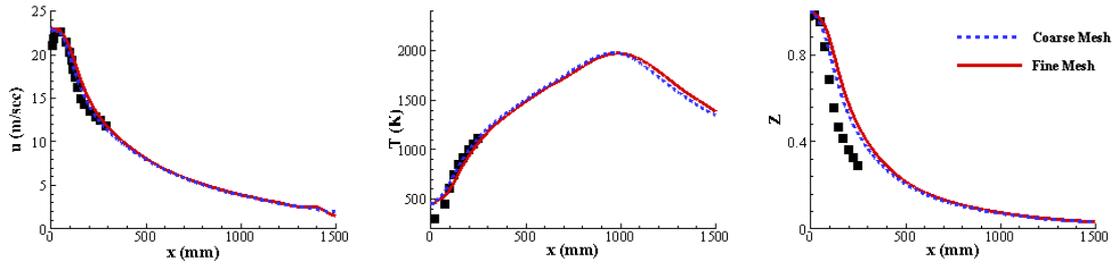

Figure 3: Centre line plots of axial velocity, temperature and mean mixture fraction using GRI 3 mechanism: solid lines are coarse mesh, dashed lines are fine mesh and symbols are measurements

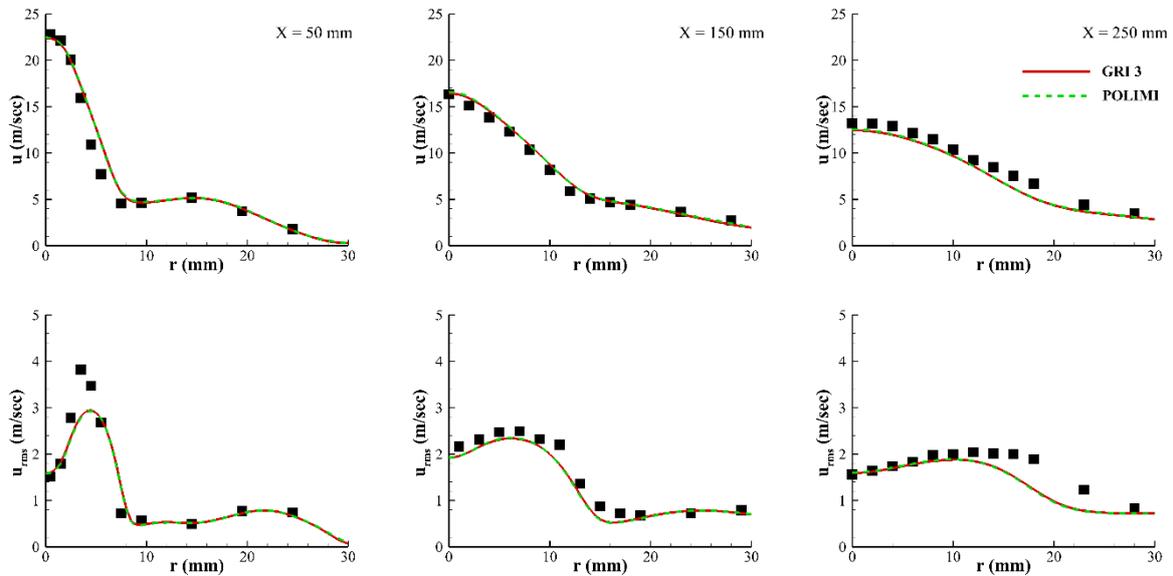

Figure 4: Radial profiles of mean and RMS axial velocity at three different locations from the fuel jet exit: lines are predictions and symbols are measurements

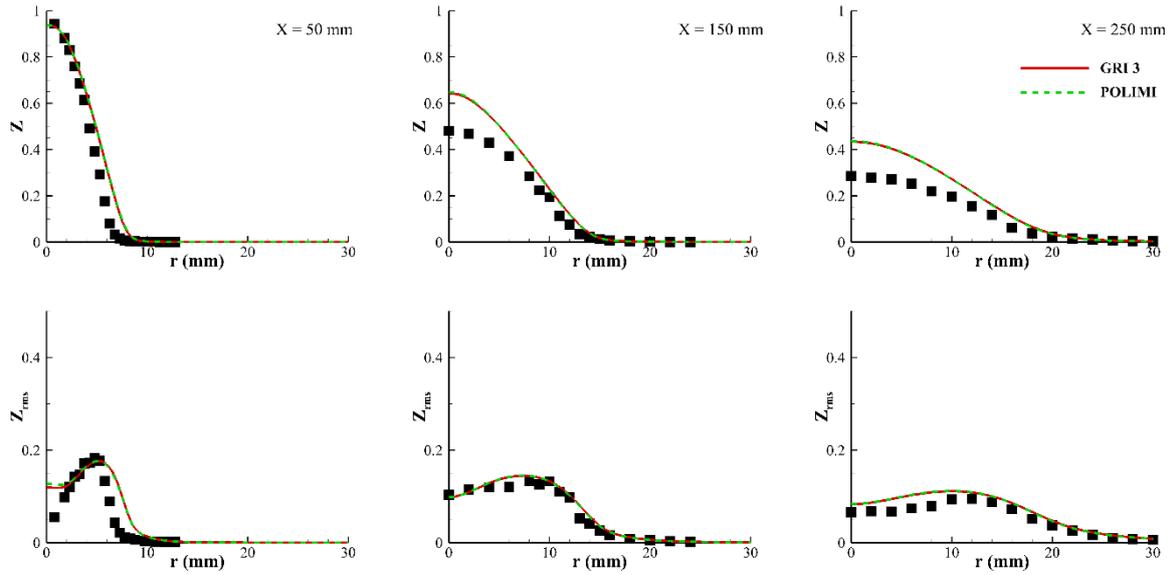

Figure 5: Radial profiles of mean and RMS mixture fraction at three different locations from the fuel jet exit: lines are predictions and symbols are measurements

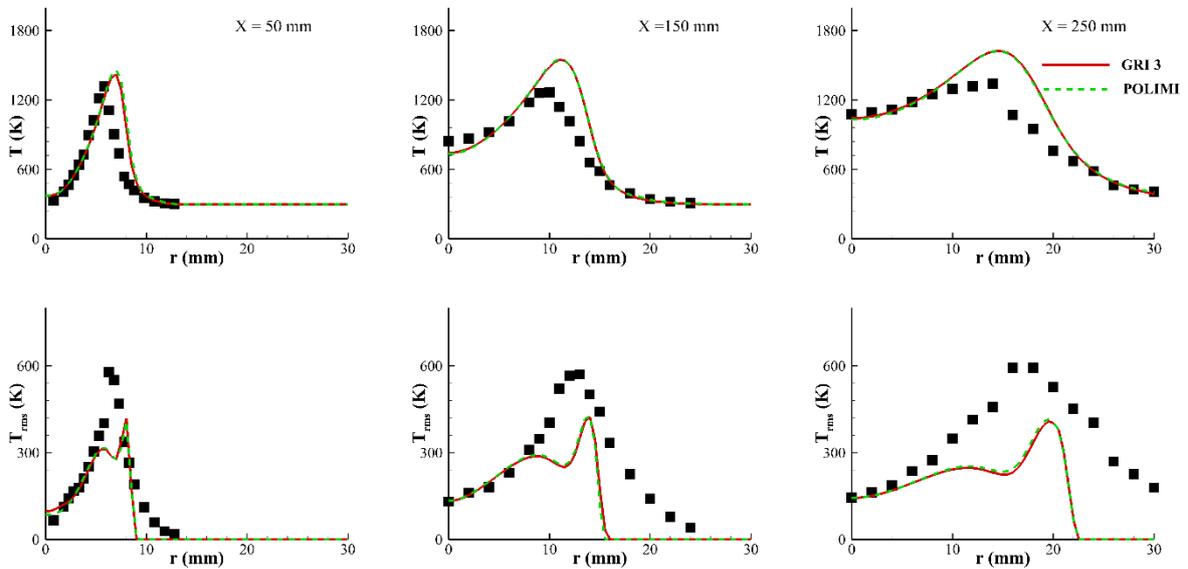

Figure 6: Radial profiles of mean and RMS temperature at three different locations from the fuel jet exit: lines are predictions and symbols are measurements

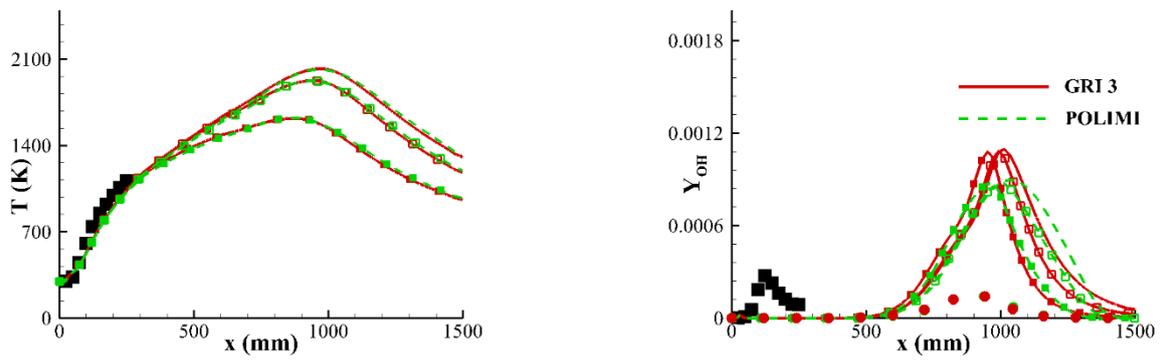

Figure 7: The centreline profile temperature and OH mass fraction. Lines without symbols are without radiation, lines with hollows symbols are with gray radiation, lines with filled symbols are with non-gray radiation, circular symbols are calculated equilibrium OH mass fraction and symbols are measurements

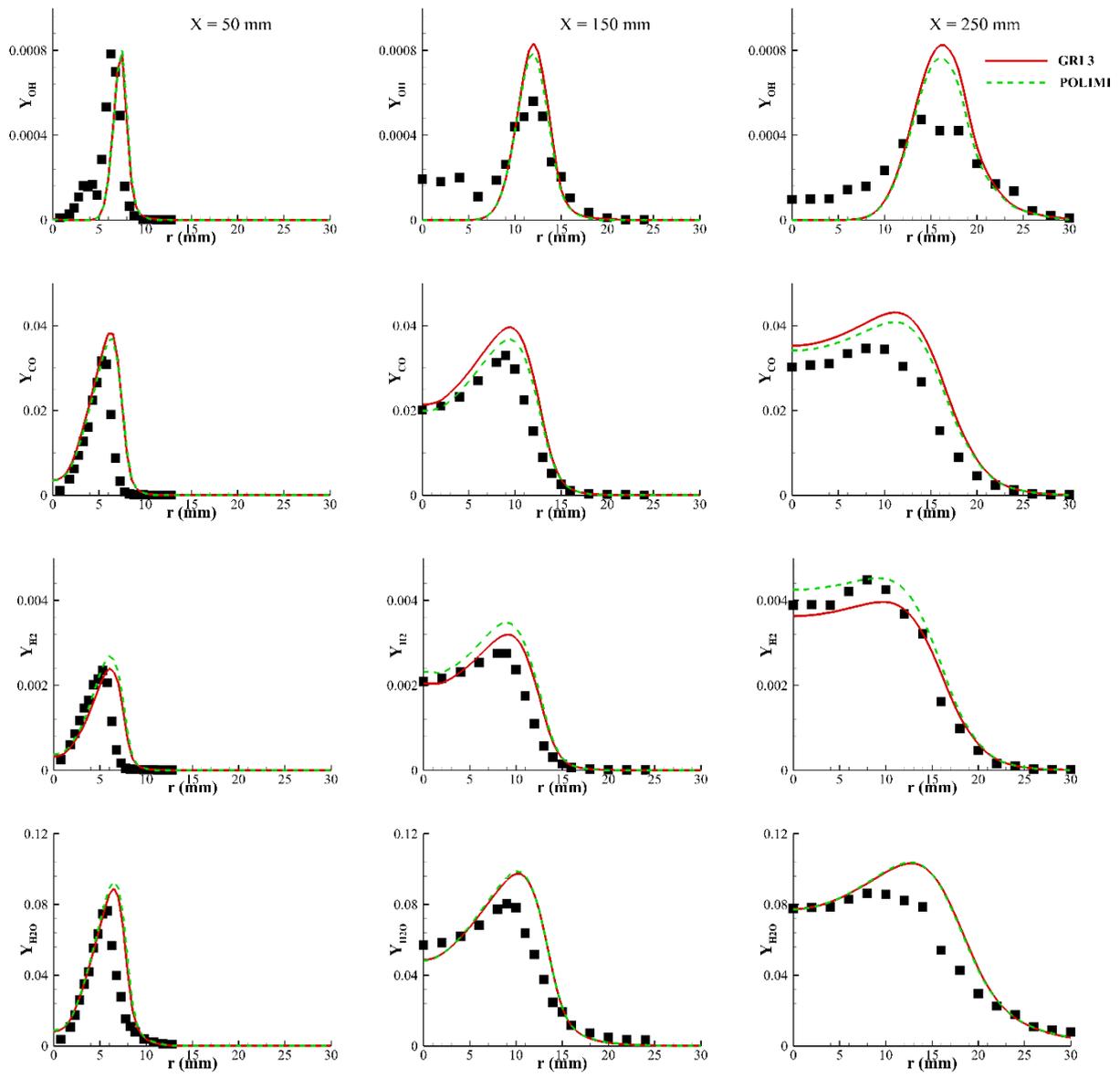

Figure 8: Radial profiles of species mass fraction at three different locations from the fuel jet exit: lines are predictions and symbols are measurements

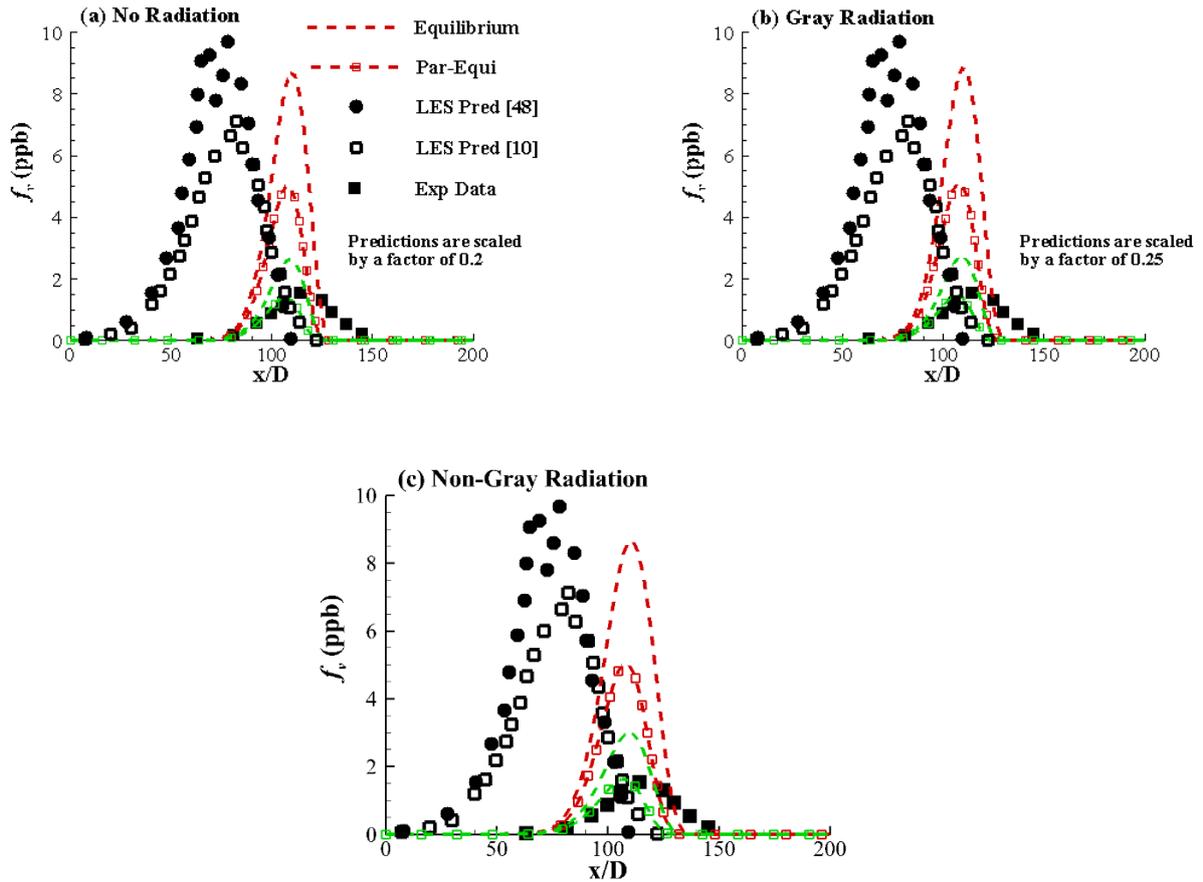

Figure 9: The centreline profile of soot volume fraction. The red line indicates GRI 3.0 mechanism. The green lines indicate POLIMI mechanism, square symbols are experimental measurements

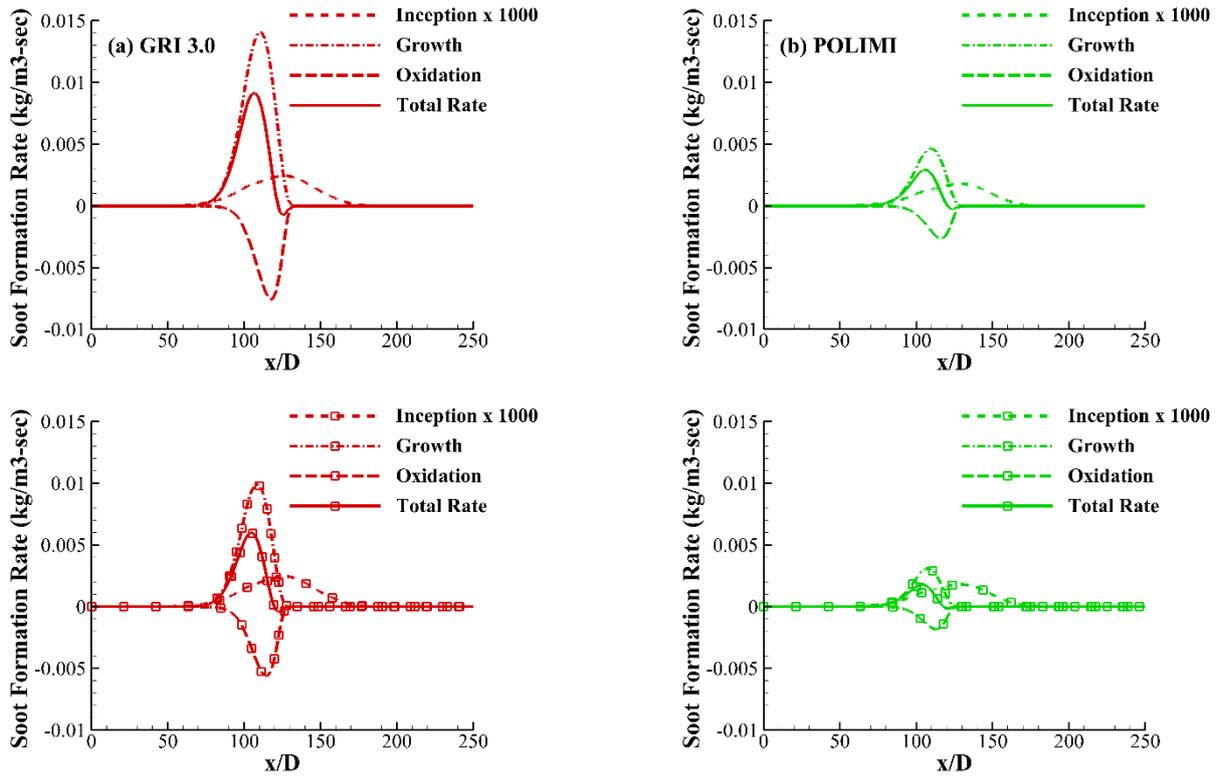

Figure 10: The centreline profiles of soot volume fraction source terms with gray radiation. The red line indicates GRI 3.0 mechanism. The green lines indicates POLIMI mechanism, Lines with symbols indicate Par-Equilibrium approach.

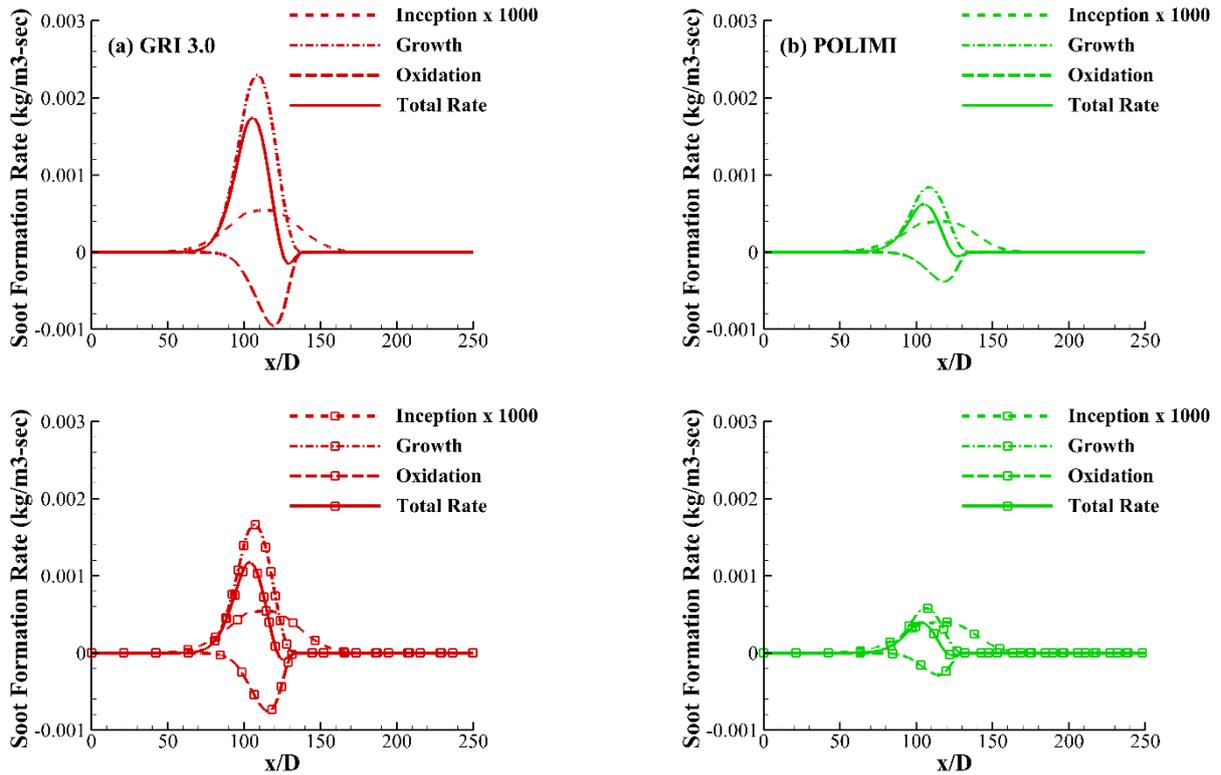

Figure 11: The centreline profile of soot volume fraction source terms with non-gray radiation. The red line indicates GRI 3.0 mechanism. The green lines indicate POLIMI mechanism, Lines with symbols indicate Par-Equilibrium approach.

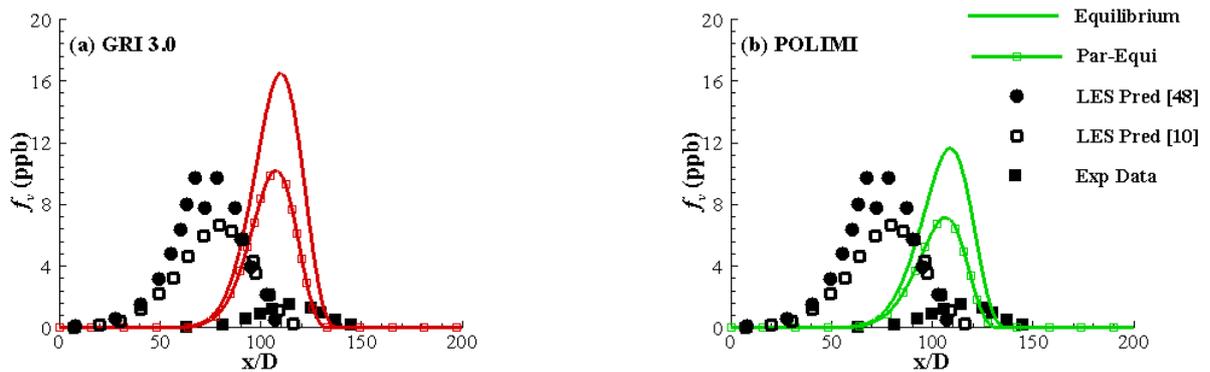

Figure 12: The centreline profile of soot volume fraction with the inclusion of non-gray radiation. The red line indicates GRI 3.0 mechanism. The green lines indicate POLIMI mechanism, symbols are measurements

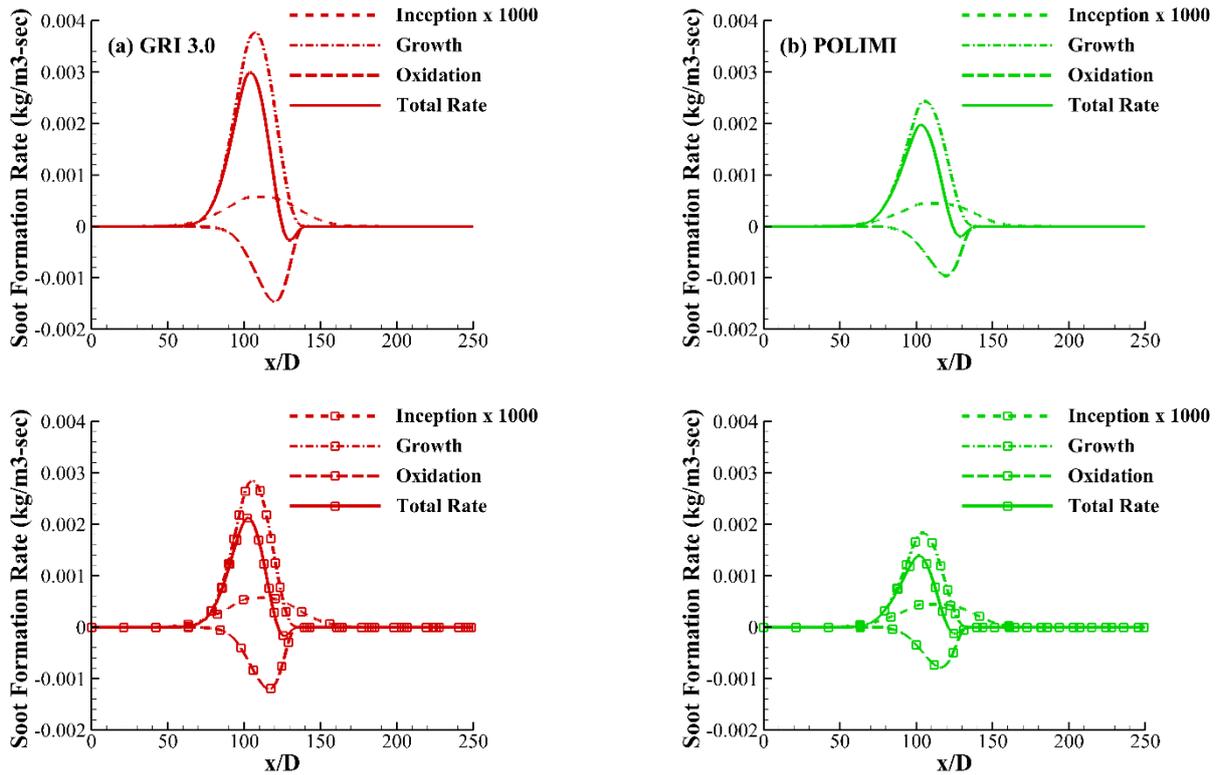

Figure 13: The centreline profile of soot volume fraction source terms with non-gray radiation and inclusion of ethylene. The red line indicates GRI 3.0 mechanism. The green lines indicate POLIMI mechanism, Lines with symbols indicate Par-Equilibrium approach.

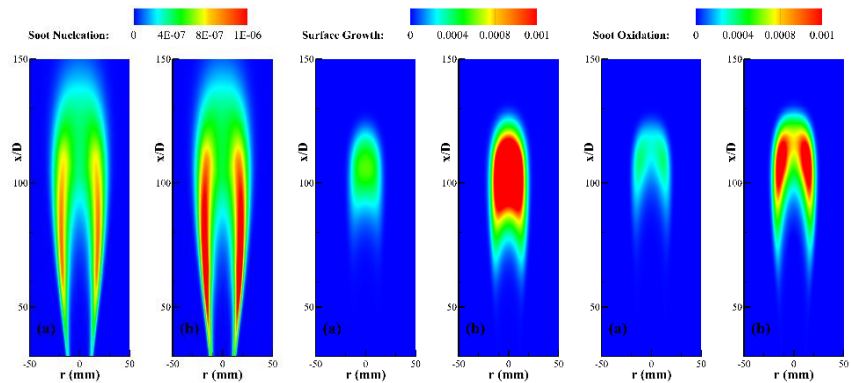

Figure 14: Contours of soot nucleation (left), soot surface growth (middle) and soot oxidation (right) using Moss Brookes model with non-gray radiation (a) With acetylene (b) with acetylene and ethylene.

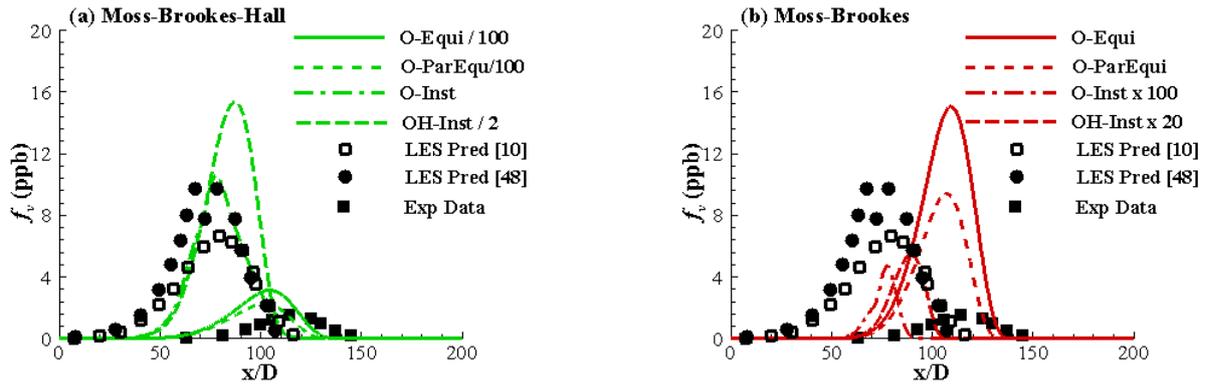

Figure 15: The centreline profile of soot volume fraction with the inclusion of acetylene, ethylene and benzene. The green line indicates GRI 3.0 mechanism. The red lines indicates POLIMI mechanism, symbols are measurements.

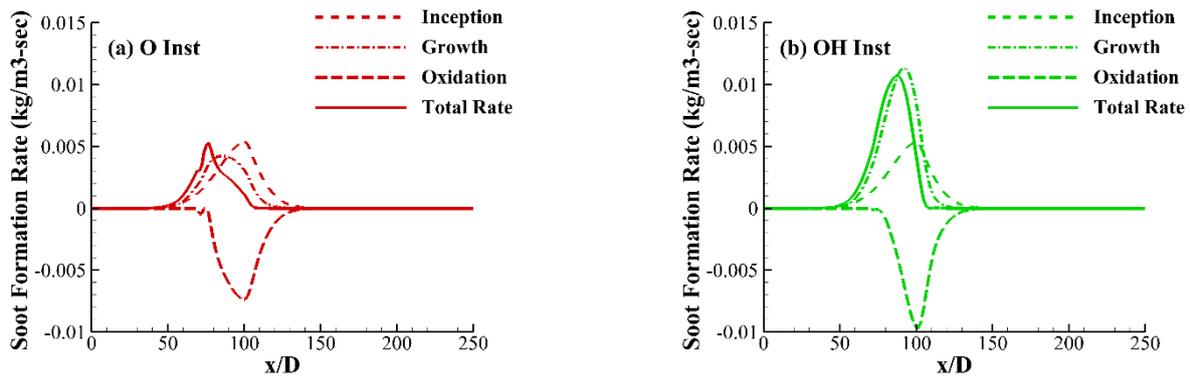

Figure 16: The centreline profile of soot volume fraction source terms with non-gray radiation and inclusion of ethylene and benzene. The red line uses O-instantaneous approach. The green lines uses OH-instantaneous approach.

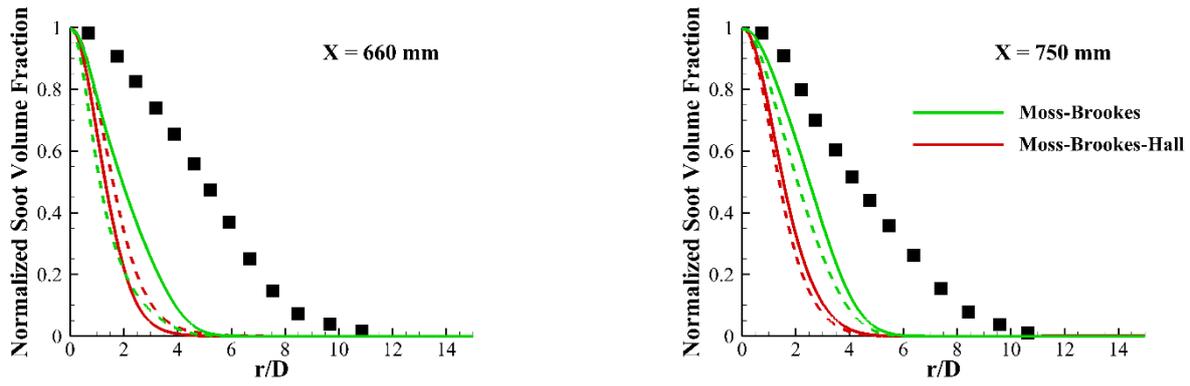

Figure 17: Radial profile of normalized soot volume fraction with soot-turbulence interactions: solid lines are O instantaneous, dashed lines are with OH- instantaneous and symbols are measurements

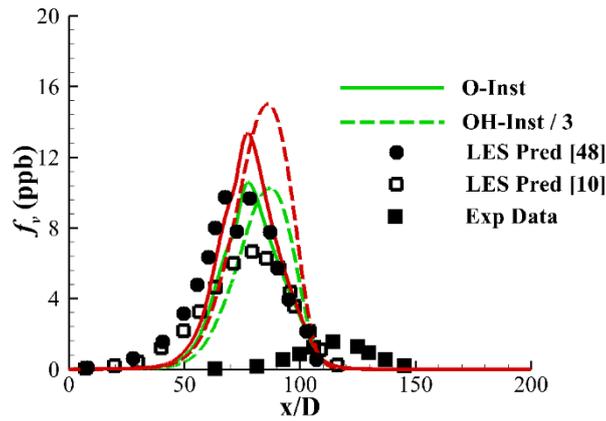

Figure 18: The centreline profile of soot volume fraction with the inclusion of acetylene, ethylene and benzene. The green line indicates without STI and the red line indicates with STI. Symbols are measurements.

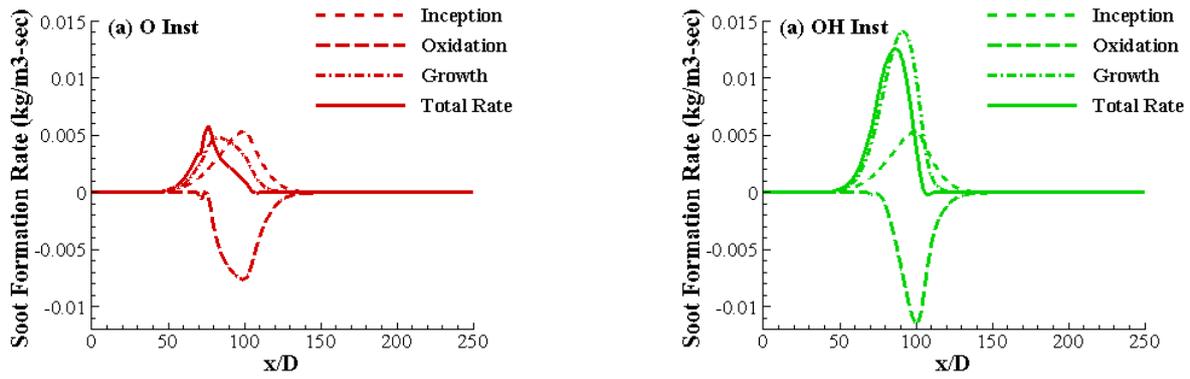

Figure 19: The centreline profile of soot volume fraction source terms with non-gray radiation and inclusion of ethylene and benzene. The red line uses O-instantaneous approach. The green lines use OH-instantaneous approach.

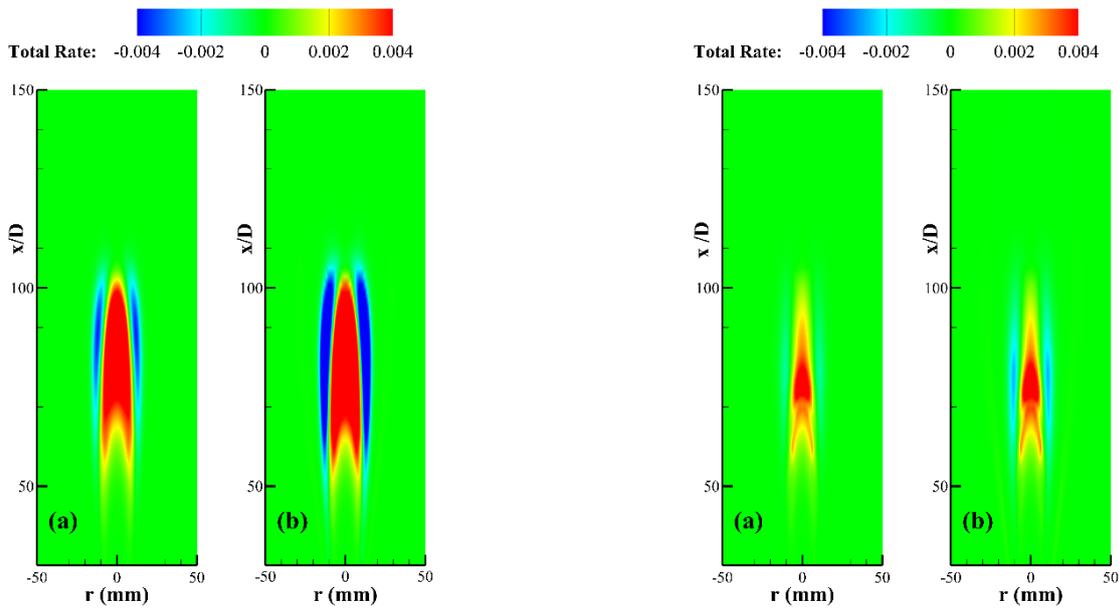

Figure 20: Contours of total soot rate using O-instantaneous approach (left), using OH-instantaneous approach (right) with non-gray radiation (a) without STI (b) with STI.

41